\documentclass[preprint,fleqn]{cas-dc}

\usepackage[authoryear,longnamesfirst]{natbib}
\usepackage{stfloats}
\usepackage{float}
\usepackage{graphicx}

\def\tsc#1{\csdef{#1}{\textsc{\lowercase{#1}}\xspace}}
\tsc{WGM}
\tsc{QE}
\begin{document}
\let\WriteBookmarks\relax
\def\floatpagepagefraction{1}
\def\textpagefraction{.001}
\shorttitle{SimPath: Mitigating Motion Sickness in In - vehicle Infotainment Systems via Driving Condition Adaptation}
\shortauthors{J.Huang et~al.}
%\begin{frontmatter}

\title [mode = title]{SimPath: Mitigating Motion Sickness in In - vehicle Infotainment Systems via Driving Condition Adaptation}                      

\author[1]{Jinghao Huang}[style=chinese,orcid=0009-0001-1498-7470]
\cormark[1]
\ead{chillaxwow@foxmail.com}
\ead{dandingxie@stu.xjtu.edu.cn}

\author[1]{Siqi Yao}[style=chinese]
\ead{cmybnysq@stu.xjtu.edu.cn}

\author[1]{Yu Zhang}[style=chinese]
\ead{zhang.yu@xjtu.edu.cn}

\affiliation[1]{ organization={School Of Mechanical Engineering},
  institution={Xi'an Jiaotong University},
  city={Xi'an},
  state={Shan xi},
  country={China}}

\cortext[cor1]{Principal corresponding author}

\begin{abstract}
The problem of Motion Sickness (MS) among passengers significantly impacts the comfort and efficiency of In-Vehicle Infotainment Systems (IVIS) use. In this study, we innovatively designed SimPath, a visual design to effectively mitigate passengers' MS and boost their efficiency of using IVIS during driving. The study focuses on the problem of irregular motion conditions frequently encountered during actual driving. To validate the efficacy of this approach, two sets of real - vehicle experiments were carried out in real driving scenarios. The results demonstrate that this approach significantly reduces passenger's MS level to a certain extent. However, due to divided attention from visual content, it does not directly improve the IVIS efficiency. In conclusion, this study offers crucial insights for the design of a more intelligent and user friendly IVIS, based on the discussion of the principle, providing strong theoretical support and practical guidance for the development of future IVIS in autonomous vehicles.
\end{abstract}

% \begin{highlights}
% \item Research highlights item 1
% \item Research highlights item 2
% \item Research highlights item 3
% \end{highlights}

\begin{keywords}
Motion Sickness \sep Visual Design \sep In - vehicle Infotainment Systems \sep Occupant
\end{keywords}

\maketitle

\begin{figure*}[h]
    \centering
    \includegraphics[width=1\linewidth]{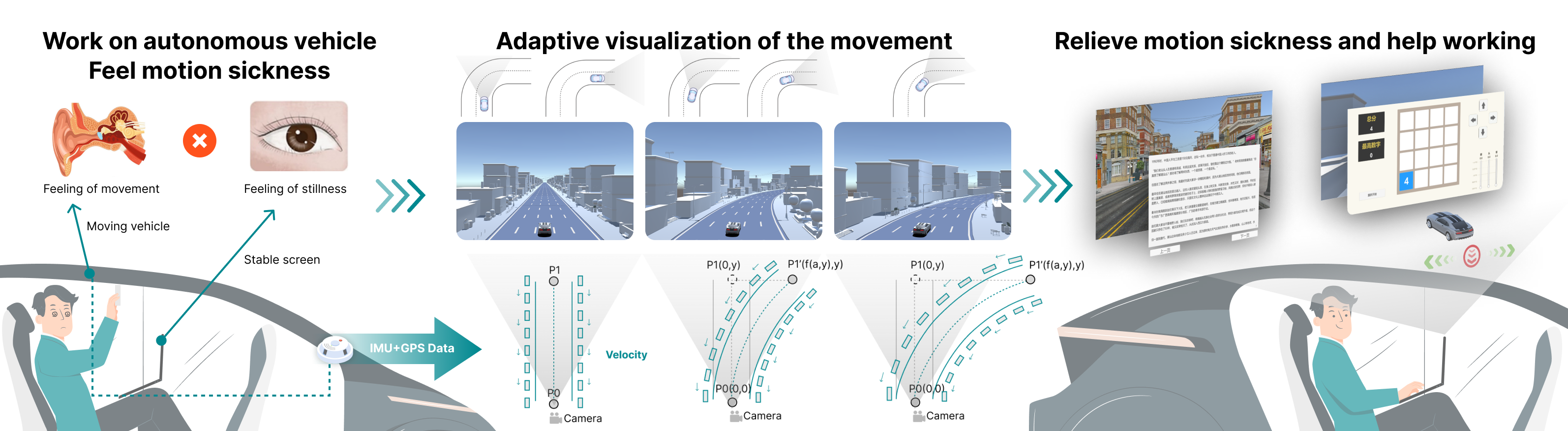}
    \caption{Graphical Abstract: Adaptive Visualization of Driving Dynamics to Mitigate Motion Sickness in IVIS}
    \label{fig:enter-label}
\end{figure*}

\section{Introduction}
As intelligent cockpit technology continues to advance, the duration passengers spend engaging with In-Vehicle Infotainment Systems (IVIS) during travel has markedly increased. Consequently, enhancing passenger comfort and efficiency in IVIS usage has become a central focus in modern IVIS design research. Motion sickness (MS), a common physiological issue encountered by passengers during travel, often arises when in motion or when motion is perceived \citep{schmidt2020international}. Typical symptoms include cold sweats, stomach discomfort, fatigue, drowsiness, headaches, nausea, and vomiting \citep{diels2016self,gallagher2018cybersickness,iskander2019car,kirst2022problem,dam2021review}, posing significant challenges to travel and work. Additionally, studies \citep{bos2004motions,colwell2008human,stroud2005preflight} suggest that alleviating MS can substantially improve IVIS usability. Thus, addressing MS is critical for enhancing comfort and efficiency in IVIS interactions, highlighting the need for urgent solutions in current IVIS design. 

The term "motion sickness" describes various nausea-related syndromes \citep{oman1990motion}. The most accepted explanation is the sensory conflict theory \citep{reason1975motion}, which posits that the brain receives conflicting neural and/or humoral signals from areas involved in spatial orientation, leading to sickness symptoms when these signals reach other brain centers \citep{iskander2019car}. In vehicles, when sensory inputs are inconsistent—specifically when vestibular motion sensations clash with visual cues or expectations based on experience—MS can occur. This theory underscores the role of sensory information and suggests that minimizing sensory conflicts could alleviate MS \citep{turner1999motion}. 

Prior studies in MS research related to conflicts between the visual and vestibular systems have investigated diverse angles, especially focusing on visual adjustment techniques. The primary concept is that human visual motion perception is primarily guided by the optic flow within images \citep{gibson1950perception}, such as delivering motion reference images\citep{mcgill2017passenger, cho2022ridevr, kuiper2018looking, pohlmann2022can, qiu2023manipulating}, addressing visual-body movement\citep{kato2008improvement, feenstra2011visual, cao2018visually}, anticipating vehicle motion\citep{sawabe2016diminished, bohrmann2022effects, li2022mitigating, diels2023design} and conveying vehicle motion state into some application\citep{apple1, apple2, meschtscherjakov2019bubble, li2025can}, that will alleviate the feeling of motion sickness. However, current research still needs a quantitative link between visual content and vehicle dynamics. Also, the usual focus is solely on MS, overlooking vision's impact on IVIS efficiency. Experiments mostly use simulators or simple routes, unlike real-world scenarios, creating a gap between research conclusions and actual outcomes. 

In conclusion, synchronizing passengers' visual and vestibular motion sensations or forewarning them of the driving route may help reduce MS. Thus, we introduce SimPath, a visual IVIS design that adaptively modifies visual content, according to a quantitative formula connecting visual content with vehicle speed and acceleration in real-world complex scenarios, to harmonize passengers' visual and vestibular motion perceptions. Additionally, an interface is crafted to deliver preemptive cues about the vehicle's driving dynamics, aiding in route prediction. This effectively lessens MS and boosts the efficiency of IVIS use in vehicles.

\section{Method}

This research examines the problem of MS caused by IVIS in moving vehicles. As numerous previous studies \citep{qiu2024acceleration, gibson1950perception} demonstrate, visual content should highlight the vestibular-detected acceleration rather than speed to alleviate MS effectively. Acceleration can be categorized into linear and steering acceleration. Consequently, SimPath, proposed in this study, leverages the Unity platform to create a virtual road environment. It adjusts the simulated road conditions based on vehicle acceleration data from the IMU and GPS, generating a visual optic flow synchronized with the vehicle's movement.

For accurate human visual speed perception in linear acceleration simulations, incorporating reference objects is essential. This study selects buildings and an actual road as reference objects, matching real world proportions. The objects' regulation is based on the linear acceleration \(a\) and motion speed \(v\) as determined by IMU and GPS sensors, allowing SimPath to synchronize the scene's motion with real-time acceleration shifts.

In simulating steering-induced acceleration, this research adopts the road-bending technique. This involves adjusting all control points on the virtual road's arc either to the left or right, creating displaced control points. The lateral deviation \(x\) of any forward road point \(P_1\) is expressed as the quadratic parabolic function \(g(a)*y^2\), where \(y\)denotes the distance from the camera \(P_0\) along the road path, and \(a\)is the steering acceleration. This formulation defines the road's bending geometry.To ensure visual display consistency, the design of function \(g(a)\) should adhere to the following rule: \(g(a)\) shows minimal variation when acceleration \(a\) is below the minimum human detectable threshold for steering or when \(a\) approaches the maximum road acceleration (typically caused by IMU sensor spikes). The precise equation is given by:

\begin{equation}  
x = g(a)*y^2 = \left\{  
\begin{array}{lcl}  %lcl-靠左对齐，ccl-居中对齐，rcl-靠右对齐
k * \sigma(z(a))*y^2 = \frac{k*y^2}{1 + e^{-z(a)}} & & {a > 0} \\
0 & & {a = 0} \\
-k * \sigma(z(a))*y^2 = \frac{-k*y^2}{1 + e^{-z(a)}} & & {a < 0} \\
\end{array}   
\right.  
\end{equation}  

\begin{equation}
z(a) = 10*	\left| a\right|-5*\frac{a_{max}+a_{min}}{a_{max}-a_{min}}
\end{equation}

\begin{figure*}[bhtp]
    \centering
    \includegraphics[width=1\linewidth]{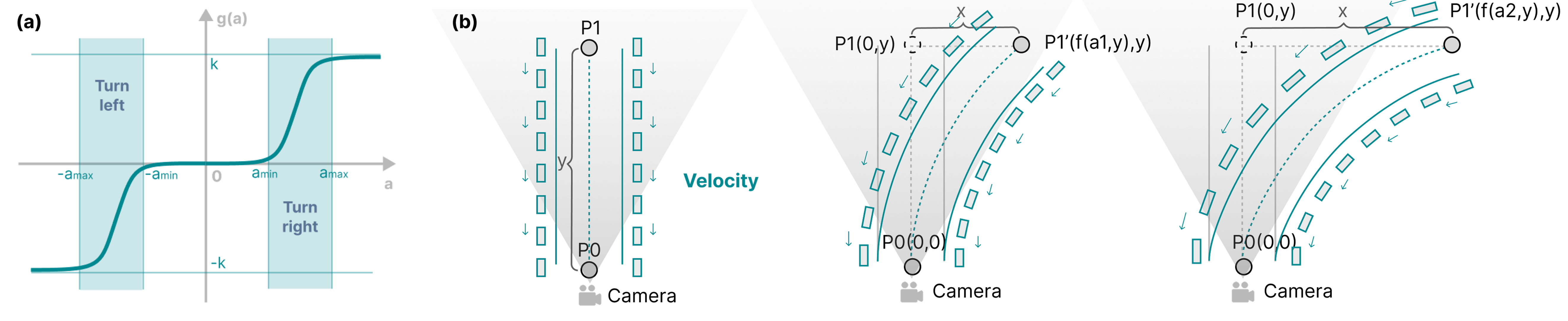}
    \caption{(a). Function graph of g(a); (b). Example of the change of point \(P1\)}
    \label{fig:enter-label}
\end{figure*}

\begin{figure*}[bp]
    \centering
    \includegraphics[width=1\linewidth]{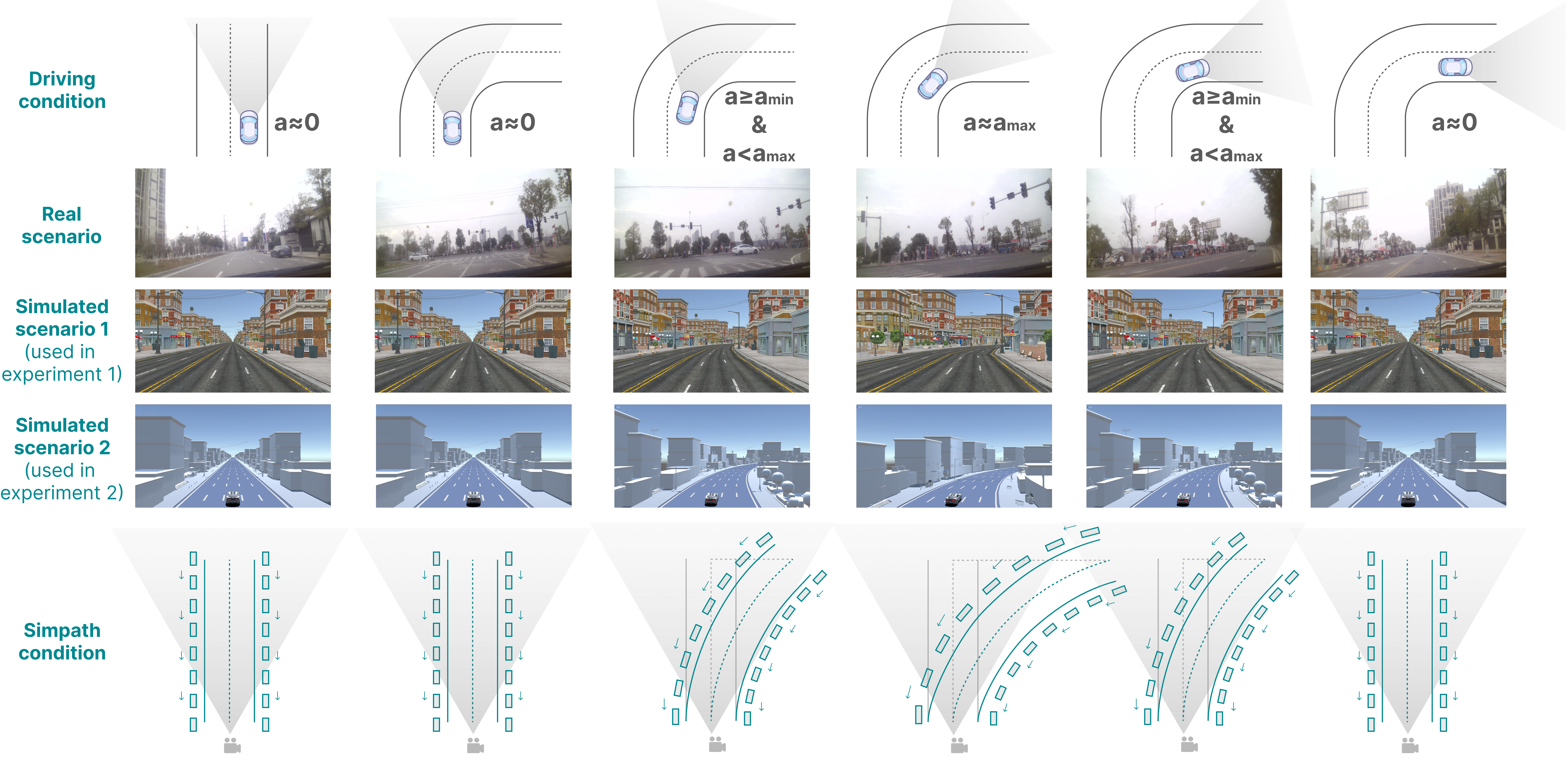}
     \caption{Images of Simpath under different driving conditions}
     \label{fig:enter-label}
\end{figure*}

Among these, \(\sigma(x)\) denotes the sigmoid function \citep{rumelhart1986learning}, which adheres to the functional requirements for \(g(a)\) . \(z(a)\) is a linear transformation equation that constrains acceleration \(a\) to \((-5, 5)\) for the \(\sigma(x)\) function. \(a_{min}\) and \(a_{max}\) are the absolute values of the minimum detectable \citep{zhiyong2016adaptive} and maximum normal steering accelerations , respectively. \(k\) is an adjustable threshold parameter. Based on pre-experiment simulations, \(k\)=0.3, \(a_{min}=2.6^{\circ}/s\) and \(a_{max}=10^{\circ}/s\) was adopted in this experiment, demonstrating great consistency with passengers' visual perception of real roads. 

As the vehicle's acceleration alters across steering phases, Figure 3 illustrates SimPath's screen states pertaining to these phases. On a curved road, during routine steering, real-time steering acceleration \(a\) adjusts. Thus, SimPath reflects varying road curvature corresponding to shifts in \(a\).

\section{Study-1-virtual scence}

\subsection{Experimental design}
This study examines 3 unique visual settings on participants on a known MS-inducing open road. The first scenario serves as a control with a blank IVIS, the second shows real-time footage from the car's front camera, and the third presents SimPath visuals reflecting both linear and steering accelerations, as depicted in Figure 4a. MS is assessed as overall and real-time experience. Passenger effectiveness using IVIS is gauged by task completion rates. In previous studies, Kennedy's 1933 Simulator Sickness Questionnaire (SSQ) is widely used to quantify MS, categorizing symptoms into eye, head, and stomach discomfort with distinct weight calculations. Participants complete the SSQ before and after the experiment (pre- and post-SSQ), while physiological indicators like ECG and Electrodermal Activity (EDA) are also monitored. For real-time MS assessment, a simplified SSQ-based Likert scale gauges eye, head, and stomach discomfort separately, and the formula in SSQ \(MS = (MS_{eye} + MS_{stomach} + MS_{head})*3.74\) is used. The scale is visibly placed at the screen's bottom-right and turns red every 30 seconds, prompting participants to update if discomfort levels change. Reading speed(seconds per page, about 1,000 Chinese characters on one page) is the chosen performance metric for the IVIS task \citep{borman1997task}.

\subsection{Experimental procedures}

\subsubsection{Subject selection}
Fifteen individuals participated in the experiment, all of whom completed the Motion Sickness Susceptibility Questionnaire (MSSQ) \citep{keshavarz2023visually} to evaluate their MS susceptibility. Participants were categorized as having medium or high levels of MS susceptibility.

\subsubsection{Experimental process}
Days before the experiment, Participants underwent an MSSQ assessment of their MS susceptibility. At the experiment's initiation, a pre-SSQ measurement was taken. Throughout the experiment, a SSQ-based Likert scale was employed to document real-time MS experiences, while physiological data were simultaneously recorded for accurate MS assessment. The experiment included a reading task displayed centrally on the IVIS screen. After completing the experiment, participants repeated the SSQ and participated in an interview by the principal experimenter to deeply explore the primary factors affecting MS changes during the study. Participants rested for at least 20 minutes after finishing one experimental scene to ensure full recovery before proceeding to the next scene. Each participant engaged in experiments across three distinct conditions.

\subsubsection{Experiment equipment and road condition}
The study was performed on an authentic open road. A high-definition external camera (2K, 30 fps) was mounted at the front driver's seat of the test vehicle, accompanied by an inertial navigation IMU + GPS sensor (Weite Intelligence WTGAHRS2). A 15.6-inch touch screen (3840x2160) was attached behind the driver's seat. To minimize visual disturbances and enhance motion sickness effects for improved data collection, a curtain obscured the window view. The computing device involved was a Mechrevo16K (RTX3060), the test vehicle was a BYD Qin PLUS, and physiological signals were registered using a Shimmer device. The experimental software environment was crafted with Unity.
The experiment involved a roughly 9.8-kilometer open road, traversed in approximately 15 minutes at an average speed of 40 km/h. A single driver ensured uniform driving style across the test. Motion Sickness Dose Value (MSDV) was evaluated to confirm the experiment's validity, detailed in Appendix A.

\begin{figure*}[h]
    \centering
    \centerline{\includegraphics[width=1\linewidth]{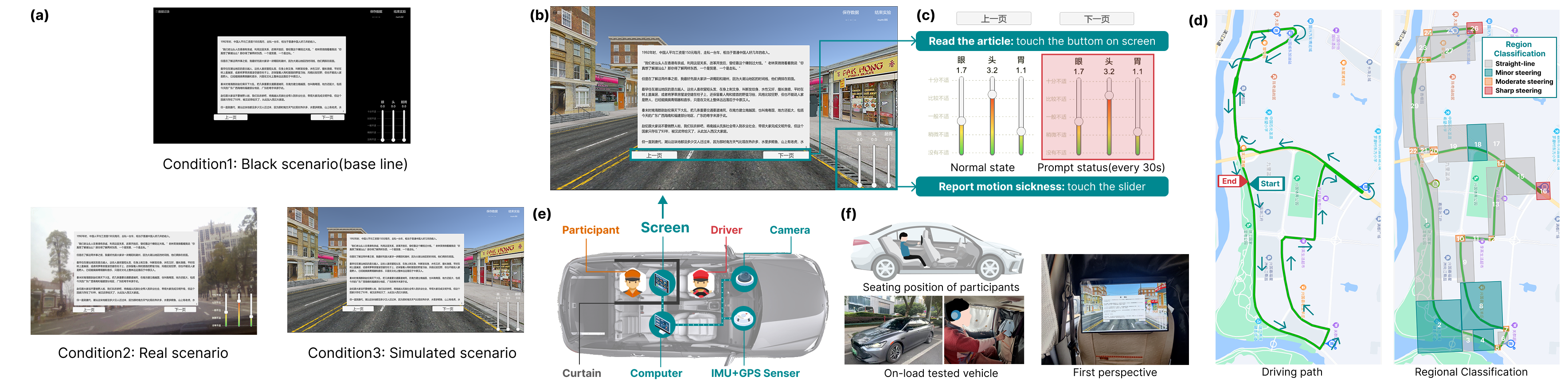}}
    \caption{(a). Display interface in three conditions; (b). Display interface in screen for participants; (c). Method for participants to read the article and report their MS value during the experiment; (d). Road condition; (e). Interior layout in the vehicle; (f). Pictures of the vehicle}
    \label{fig:enter-label}
\end{figure*}

\subsection{Experimental Results and Analysis}
\subsubsection{MS Level Analysis}
Pre-MS and Post-MS levels in this study was assessed using Pre-SSQ and Post-SSQ administered prior to and following the experiment(Figure 5a), measuring passengers' MS levels before and after the vehicle ride. Analysis of variance results show significant differences in the three IVIS conditions: condition1 (F(2,42)=12.886, p = 0.001**, \(\eta^{2}\) = 0.315), condition2 (F(2,42)=8.583, p = 0.007**, \(\eta^{2}\) = 0.235), and condition3 (F(2,42)=12.3, p = 0.002**, \(\eta^{2}\) = 0.305), indicating notable changes from pre-SSQ to post-SSQ. This suggests that the experimental procedures significantly influenced the participants' MS.  However, no significant differences in post-SSQ values were found across different scenes (F(2,42)=1.986, p = 0.15, \(\eta^{2}\) = 0.086), suggesting the lack of impact by scene variations on MS post-journey. 

Real-time MS levels were recorded through self-reports at intervals of up to 30 seconds(Figure 5b). ANOVA results indicate significant differences among IVIS conditions (F(2,42)=3.876, p = 0.029*, \(\eta^{2}\) = 0.156), with subsequent comparisons revealing that average MS in condition1 (M = 14.32, SD = 6.54) was significantly higher (p = 0.008**) than in condition2 (M = 9.17, SD = 3.66).

\subsubsection{Electrocardiogram and Electrodermal Activity}
The experiment utilized ECG parameters, such as average heart rate, RR interval, and LF/HF ratio, alongside EDA measures like average skin conductance level(SCL), to evaluate MS levels(Figure 5d). Variance analysis revealed no significant differences across conditions in heart rate (F(2,36)=2.003, p = 0.148, \(\eta^{2}\) = 0.087), RR interval (F(2,36)=2.142, p = 0.13, \(\eta^{2}\) = 0.093), LF/HF (F(2,36)=1.134, p = 0.331, \(\eta^{2}\) = 0.051), or Skin Conductance Level (SCL) (F(2,36)=0.783, p = 0.464, \(\eta^{2}\) = 0.036).

\subsubsection{Task Performance Analysis}
In this study, reading speed (seconds per page) served as a measure of task performance(Figure 5c). An analysis of variance revealed no significant variation in reading speed across different IVIS conditions (F(2,14)=0.645, p = 0.53, \(\eta^{2}\) = 0.03). The Pearson correlation test also found no significant relationship between reading speed and MS levels (r = 0.086, p = 0.575), indicating that reading speed is not influenced by visual content.

\begin{figure*}[h]
    \centering
    \includegraphics[width=1\linewidth]{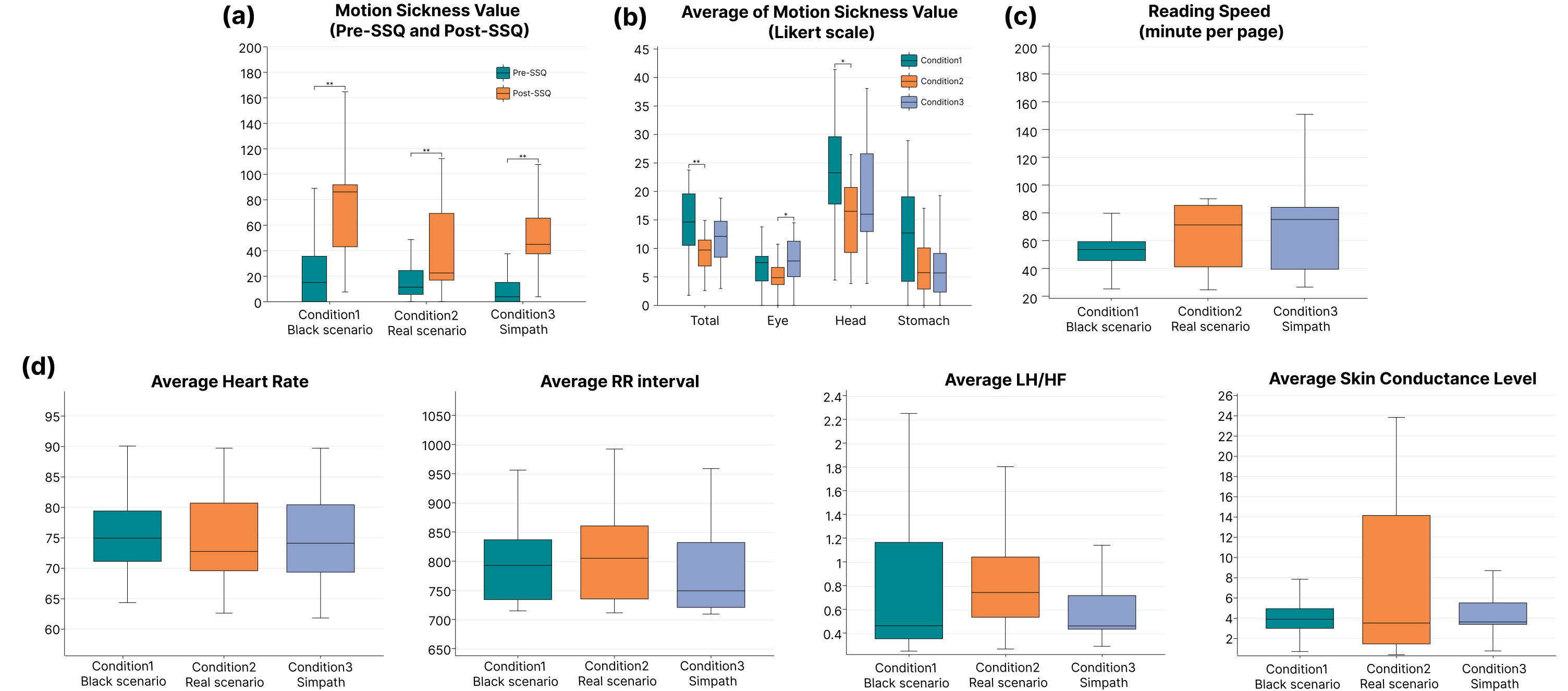}
    \caption{(a). MS Value across the three conditions (Pre-SSQ and Post-SSQ); (b). Average MS value(Likert scale) across the three conditions; (c). Reading speed across the three conditions;  (d). Average HR, RR interval, LF/HF and SCL across the three conditions;  Significance levels are indicated as *p < 0.05, **p < 0.01, and ***p < 0.001; All box plots in this article was drawn using ChiPlot (https://www.chiplot.online/)}
    \label{fig:enter-label}
\end{figure*}

\subsection{Discussion}
\subsubsection{MS Sensations}
In Experiment 1, the level of MS among passengers significantly decreased under condition2 when real-driving footage was shown, as it effectively modified visual motion perception. Conversely, condition3 with SimPath did not achieve a notable reduction in MS, potentially due to design flaws. Firstly, SimPath's control may suffer from errors in synchronizing vehicle and visual motion, as it lacks a camera's real-time feedback and has a latency of around 40ms, diminishing its anti-MS effect. Secondly, using a first-person perspective with SimPath results in screen edge object motion appearing faster than the vehicle's actual speed, misaligning visual and actual motions—this might be improved with a third-person perspective. Lastly, condition3 caused significant eye strain(M = 7.86, SD = 4.03), which is significantly higher(p = 0.034*) than condition2(M = 5.00, SD = 2.67), as noted by six participants in subjective interviews, who found SimPath's color settings too vivid, resulting in eye discomfort. Although an enhanced street model aimed to increase virtual-real similarity, its intense colors negatively impacted eye comfort, complicating eye-discomfort evaluations in the SSQ and affecting MS severity assessment. Thus, subsequent experiments should prioritize appropriate IVIS color selection, avoiding excessive saturation and favoring colors that are gentle on the eyes to minimize experimental interference \citep{li2022mitigating,elliot2014color}.

\subsubsection{Task Performance}
Experiment 1 revealed no meaningful differences in reading speeds across various IVIS, with excessive variance being unsatisfactory. This may relate to our choice of task to assess IVIS efficiency. Interviews after the experiment showed diverse participant reactions; three noted no link between reading speed and MS levels, eight read slower due to dizziness, while four read faster but with poor comprehension. These findings suggest reading speed varies during MS(r = 0.086, p = 0.575), undermining the task's effectiveness in gauging IVIS performance without content-specific questions.

\subsubsection{Physiological Responses}
In Experiment 1, ECG and EDA signals were utilized to evaluate MS levels, although outcomes were unsatisfactory. "No significant differences in ECG and EDA data were found across experimental conditions," aligning with similar research findings \citep{paredes2018driving, elsharkawy2024sync}. Nonetheless, heart rate slightly declined in condition 2 (M = 75.77, SD = 7.92) and condition 3 (M = 75.63, SD = 8.22) compared to condition 1 (M = 77.51, SD = 7.71). Several factors might account for the absence of significant SCL data differences. First, equipment-related issues during the experiment, such as temperature and humidity fluctuations, could have impacted EDA signal collection, alongside unforeseen events causing data instability \citep{kettner2017towards}, hindering accurate signal acquisition. Second, concerns exist that physiological signals may inaccurately reflect MS level \citep{shupak2006motion}. While ECG and EDA signals indicate certain physiological changes, their sensitivity and specificity for MS-related reactions might be insufficient. Thus, relying exclusively on ECG and EDA signals for MS level measurement has limitations. Future studies might incorporate specific physiological indicators, such as electrogastrogram \citep{cowings2000autogenic}, to improve accuracy in assessing MS levels. Although sensory-input alignment was anticipated to clearly affect EDA data, this experiment didn't show such an effect.

\subsubsection{Measurement Methods of MS Sensations}
In this study, post-SSQ values showed no notable differences among various IVIS conditions. Contrastingly, the simplified SSQ-based Likert scale revealed significant disparities. Furthermore, a strong correlation was found between post-SSQ and Likert scale values (r = 0.546, p = 0.001**). Consequently, the Likert scale demonstrated superior measurement accuracy for MS levels compared to both pre- and post-SSQ, likely due to the fluctuating nature of MS sensations. The SSQ's inability to capture real-time data results in decreased precision, while the Likert scale effectively records immediate MS experiences. Future studies should exclusively utilize the  simplified SSQ-based Likert scale for assessment.

\begin{figure*}[h]
    \centering
    \includegraphics[width=1\linewidth]{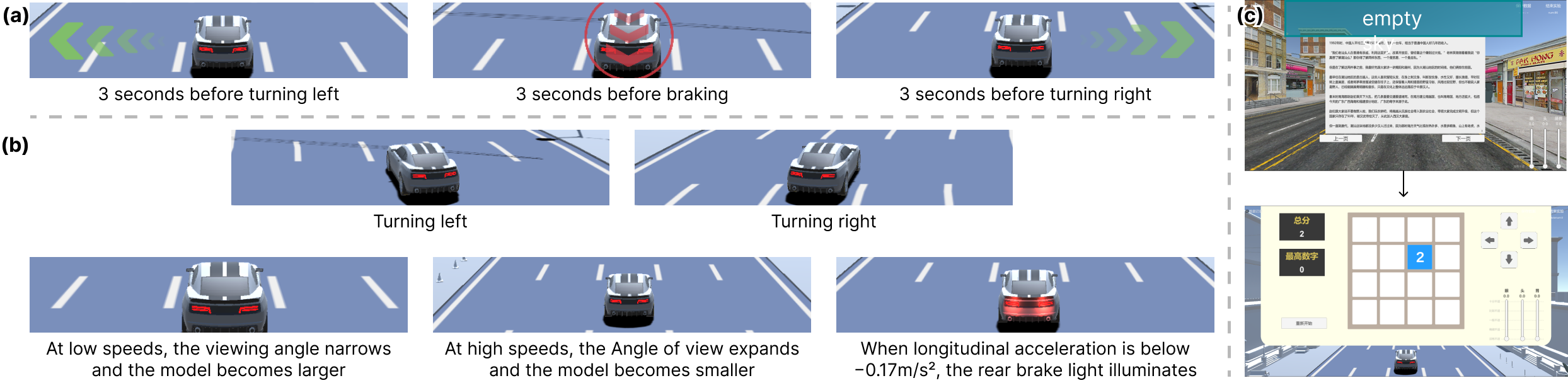}
    \caption{Changes of the method: (a). difference of the display interface in screen for participants; (b). New symbol and the car model}
    \label{fig:enter-label}
\end{figure*}

\section{Study-2-virtual scence and motion prdication}
\subsection{Changes in Experimental Methods}
The preceding chapter revealed that SimPath did not significantly lower MS levels. Prior studies suggest that pre-emptively informing participants of a vehicle's path can mitigate MS \citep{diels2016self, rolnick1991driver}. To optimize Simpath, this chapter introduces a motion trajectory prompt, enabling participants to foresee vehicle movement, thus potentially reducing MS. The presentation of linear and steering accelerations remains consistent, as depicted in Figure 6.

\subsubsection{Addition of Symbol (Prompt)}
Integrating anticipatory prompts into the virtual image is a crucial strategy for minimizing MS. This research implements anticipatory prompts through a symbol positioned in the screen's lower segment, as illustrated in the Figure 6a. Since this area avoids the primary optical flow region, it does not disrupt the optical flow display and can function as a prompt without using the main display space. In the experiment's setup, GPS locates the vehicle. When the vehicle approaches a turn or deceleration zone,  the symbol activates 3 seconds prior as a prompt, flashes at 1s intervals, and extinguishes 1s after steering concludes.

\subsubsection{Addition of Car Model}
This study refines the IVIS display by shifting from a first-person to a third-person perspective and adding a car object as a situational cue to more accurately depict the driving state(Figure 6b). This adjustment seeks to enhance the participants' visual perception of motion. Additionally, when the vehicle slows down, the rear brake lights are activated, aligning the simulation more closely with actual driving scenarios.

\subsubsection{Modification of the Overall Task Interface}
As the interface's upper section predominantly showcases the static background sky, which produces negligible optical flow and offers limited visual motion cues, the task interface is extended to encompass this region. Consistent with the covert attention, which aids us in monitoring the environment \citep{carrasco2011visual}, the car model and symbol are positioned in the covert attention area below the screen,  for peripheral awareness of information that is not the main focus\citep{grudin2001partitioning,fahle1991motion}, as illustrated in the Figure 6c.

\subsection{Experimental Design}
This study examines 4 unique visual settings on participants on a known MS-inducing open road. In the first, no dynamic visuals appear; the second shows a symbol at the screen's base; the third presents only SimPath; and the fourth features both SimPath and the symbol, as depicted in Figure 6. MS levels is measured in real-time using a simplified SSQ-based Likert scale derived from the SSQ, adhering to the methodology from experiment 1. Given the results from experiment 1, the reading task is deemed inappropriate for performance assessment in this setting. Therefore, the study employs the 2048 game as the IVIS task due to its relevance to participants' concentration and effort, with the game displayed on the main screen. A gamepad is supplied for ease of use. Details of operation are shown in Figure 7a and 7b.

\begin{figure*}[h]
    \centering
    \includegraphics[width=1\linewidth]{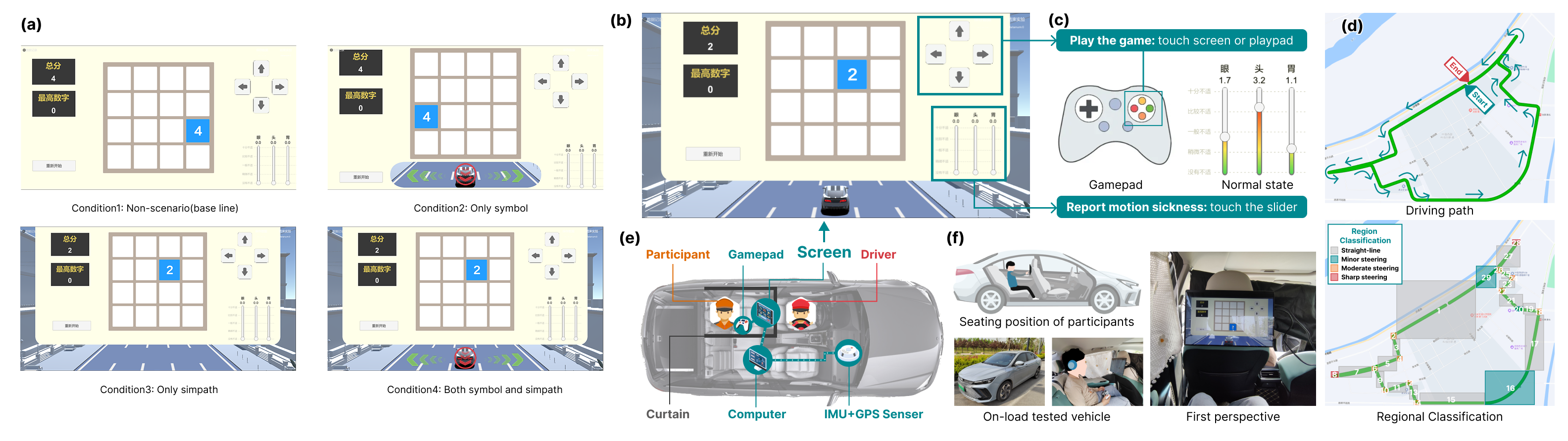}
    \caption{ (a). Display interface in four conditions; (b). Display interface in screen for participants; (c). Method for participants to play the game and report their MS value during the experiment; (d). Road condition; (e). Interior layout in the vehicle; (f). Pictures of the vehicle}
    \label{fig:enter-label}
\end{figure*}

\subsection{Experimental Procedures}
Ten individuals participated in the experiment, all of whom completed the MSSQ \citep{keshavarz2023visually} to evaluate their MS susceptibility. Participants were categorized as having medium or high levels of MS susceptibility.
The experimental methodology and in-car configuration are almost the same as in Experiment 1, with the substitution of the vehicle by a Geely Emgrand L, maintaining the same configuration as the vehicle utilized in Experiment 1, and a similar road has been selected due to traffic restrictions.
\subsection{Experimental Results and Analysis}
\subsubsection{MS level}
In this study, real-time MS levels were assessed through subjective reports submitted every 30 seconds(Figure 8a). ANOVA results indicated significant variation across visual conditions (F(3, 36)=3.174, p = 0.041*, \(\eta^{2}\) = 0.051). Post-hoc analyses found that condition1 (M = 1335.8, SD = 399.713) yielded a significantly higher MS level (p = 0.006**) than condition4 (M = 12.58, SD = 3.82), and  was also significantly higher (p = 0.046*) compared to condition2 (M = 14.13, SD = 4.07). This suggests that the SimPath interface with symbol prompts (condition4) notably affects the MS level, whereas IVIS with visual symbols (condition2) and SimPath (condition3) tend to mitigate it.

\subsubsection{Task Performance Analysis}
In this study, the aggregate game score from a single trial was employed to evaluate task performance(Figure 8b). ANOVA results indicated no substantial differences in scores across distinct visual settings (F(3, 36)=3.174, p = 0.036*, \(\eta^{2}\) = 0.115). Subsequent post-hoc analyses found that condition3 (M = 1560.6, SD = 216.608) had a significantly higher score than condition2 (M = 1245.6, SD = 237.309, p = 0.011*) and condition4 (M = 1235.8, SD = 155.653, p=0.013*). Additionally, Pearson correlation analysis showed no significant linkage between the game score and MS level (r = 0.201, p = 0.214). The findings imply that the SimPath interface, devoid of symbol prompts in condition3, markedly aids passengers in achieving superior task performance.

\begin{figure}[h]
    \centering
    \includegraphics[width=1\linewidth]{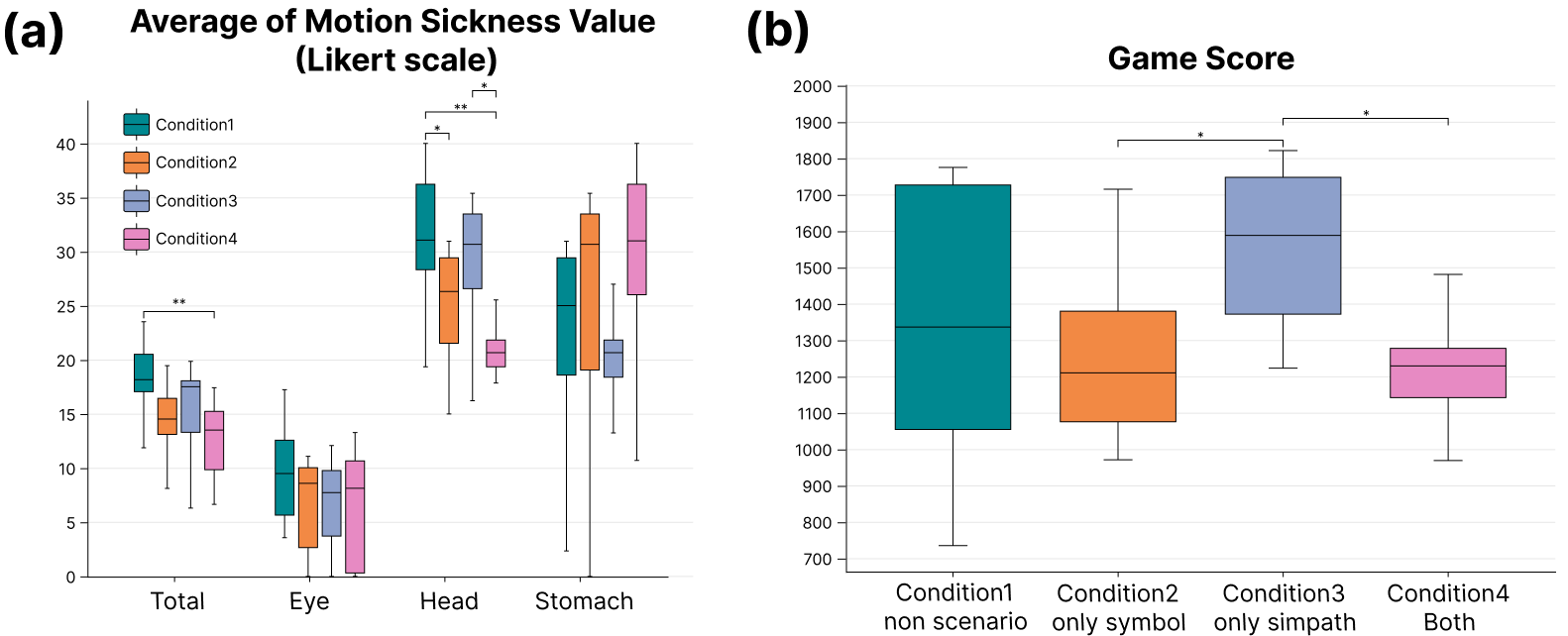}
    \caption{Data analysis: (a). Average MS value(during each condition) across the four conditions; (b). Game score across the four conditions; Significance levels are indicated as *p < 0.05, **p < 0.01, and ***p < 0.001;}
    \label{fig:enter-label}
\end{figure}

\subsection{Discussion}
\subsubsection{MS level}
Experiment 2 finds that, in condition 4, passengers experience significantly lower MS levels compared to the other three conditions. This suggests that IVIS with symbols and SimPath is highly effective in reducing MS. The primary reasons are that symbol-based car models help decrease MS by enhancing motion perception during vehicle movements, allowing for a better correlation with vestibular cues, thus reducing MS. Additionally, symbols provide a forecast for turns and decelerations, enabling participants to predict and minimize vestibular impacts, reducing MS symptoms. However, effectiveness depends on noticing these symbols. In condition 2, the lack of SimPath made it difficult for three subjects to detect situational changes due to small interface size, which could not capture attention. Although SimPath improves motion perception, alone it does not significantly lower MS. Five participants noted that optical flow mitigates sensory conflict between vision and vestibular system, easing MS, while three overlooked SimPath changes due to low color saturation. In summary, IVIS with SimPath and symbols effectively aids participants in comprehending vehicle conditions, lowering MS levels.

\subsubsection{Measurement Methods of MS Sensations}
Experiment 2 demonstrated a notably high game score in condition 3, indicating that the IVIS symbol directs participants' focus toward the vehicle's motion and environmental changes, thereby diverting attention away from the IVIS. With the symbol's removal, attention shifts back to the game, improving scores. This symbol presence significantly affects attentional focus and reduces IVIS efficiency. Other results showed no notable link between IVIS efficiency and MS level (r = 0.018, p = 0.917), as participants prioritized the car model under MS, highlighting that distraction affects efficiency more than MS. Incorporating visual trajectories to mitigate MS will decrease efficiency due to distraction. Future product design must balance IVIS efficiency and MS reduction.

\section{Discussion and Limitation}
\subsection{Design approach and theoretical conjecture}
Informed by sensory conflict theory, the Simpath is designed to mitigate rear-seat passengers' motion sickness by quantifying the effects of vehicle dynamics on IVIS images. Notably, while rich in empirical evidence, the study lacks foundational theoretical depth, particularly in its handling of the steering function g(a), the vehicle model, and the design prompts within the practical implementation phase. All these aspects focus on the alignment between interface design and actual driving environments, alongside visual aesthetics. Consequently, the study aims to deepen the theoretical basis underpinning Simpath.

Recently studies\citep{liu2022motion, wada2024effects,emond2025can} suggests that motion sickness primarily stems from a conflict in perceiving the visual vertical. Within a vehicle's non-inertial frame during steering, centrifugal force is a fictitious force describing the "centrifugal tendency" passengers perceive due to inertia. Thus, from the perspective of our analysis, the resultant force of centrifugal force and gravity alters the vestibular system's vertical plane perception, resulting in an angle deviation, $\alpha$. Drivers, who constantly process visual data from the road and anticipate vehicle dynamics, adjust their posture proactively; thus, their active tilt angle correlates with steering acceleration \citep{mestre2012we, zikovitz1999head}, creating an offset angle $\beta$ between the visual and absolute vertical line. Aligning $\alpha$ and $\beta$ mitigates perceptual conflict, reducing motion sickness for drivers, unlike passengers who lack dynamic visual feedback and anticipation from road, exacerbating their symptoms, as illustrated in the Figure 9.

\begin{figure}[h]
    \centering
    \includegraphics[width=1\linewidth]{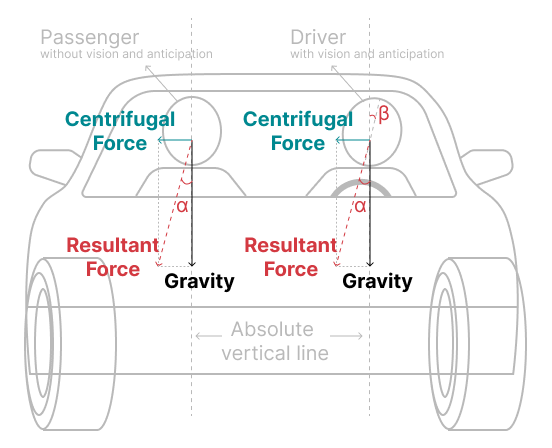}
    \caption{Force and state analysis diagram of drivers and passengers during vehicle turning (vehicle as reference frame)}
    \label{fig:enter-label}
\end{figure}

Thus, to mitigate visual-vestibular mismatch, passengers can be instructed to slightly tilt their heads. The Simpath employs both physiological and psychological methods to achieve this. Physically, it prompts head tilting by altering virtual road line intersections on IVIS, as illustrated in the Figure 10, aligning with drivers' instinctive gaze fixation on the turning path \citep{mestre2012we, zikovitz1999head}, and mitigating motion sickness as per studies on screen rotation \citep{kato2008improvement, feenstra2011visual, hwang2022adaptive}. Psychologically, it generates a steering prompt scene resembling actual roads, activating passengers' steering memories, aligning with findings \citep{li2022mitigating, reuten2023effectiveness}. These combined mechanisms theoretically encourage passengers to promote inward head tilting during turns, minimizing visual-vestibular conflicts in vertical plane perception. However, head movements were not quantified, highlighting a study limitation.

\begin{figure}[h]
    \centering
    \includegraphics[width=1\linewidth]{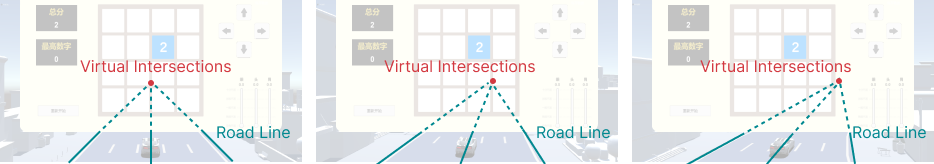}
    \caption{Virtual intersection on IVIS in three conditions}
    \label{fig:enter-label}
\end{figure}

Building on these analysis, future works will focus on: 1) using accurate mathematical modeling to transform vertical plane tilt-induced spatial alterations into 2D display parameters, facilitating IVIS interface design ; 2) employing a head-mounted IMU to track passenger head movements in response to visual stimuli. This will advance the understanding of visual design in mitigating motion sickness and offer theoretical and technical guidance for developing new low-motion-sickness IVIS systems.

\subsection{Limitations in experiment}
Experimental environment: this study utilized a real open road to conduct experiments, confirmed by the precise calculation of MSDV, demonstrating that real driving does not affect experimental reliability, thus ensuring result validity. Details are in Appendix B. Furthermore, with an increase in IVIS passenger MS experience data, deep learning can be used to analyze MS sensation traits, facilitating the creation of a prediction model for passenger MS sensations in vehicles using IVIS.

Sample and road condition: With a small sample size and limited road condition diversity, the experiment's result validity was compromised. Future research should incorporate more quantitative, canonical road scenarios and recruit additional participants to validate the feasibility of Simpath.

Task performance assessment: At First, the chosen tasks (reading/games) diverge from classic IVIS tasks, and the feasibility was not rigorously tested, compromising external validity. Secondly, attention was not measured, rendering it impossible to quantify SimPath's impact on attentional engagement. Hence, future studies should validate the feasibility of non - typical tasks and employ quantitative indicators, like eye-tracking metrics, to assess attention distribution, using the Area of Interest (AOI) display area data for comprehensive analysis of IVIS impacts on attention. 

\section{Conclusion}
This study presents SimPath, a novel method for dynamically modulating IVIS displays according to vehicle acceleration, especially in erratic driving situations. The main goal is to reduce MS in passengers engaging with IVIS while optimizing its functional performance. Experiments show that using SimPath with symbols significantly alleviates passenger MS, though it does not enhance IVIS efficiency due to visual content distraction. Thus, while motion reference images can enhance passenger comfort, they do not boost task efficiency, such as office work or entertainment.

Moreover, this study explores the mechanisms by which steering displays alleviate motion sickness, suggesting the integration of SimPath into various IVIS applications like gaming, remote meetings, and media use for improved outcomes. This integration aims to decrease motion sickness while balancing passenger engagement, driving safety, and content interaction (such as videos, meetings, and chats) that might impact external vision. Effective implementation requires software engineers to examine the distribution of passenger attention and its effects on external vision, thereby optimizing IVIS for varied content types. Such an approach enhances the understanding of passenger requirements, pushing IVIS design toward a user-focused and efficient model, offering substantial theoretical and practical foundations for future advancements in autonomous vehicle IVIS.

\section*{Essential Experimental Note}
This preprint focuses on alleviating motion sickness through visual cues and investigating the associated costs in practical applications (e.g., reduced performance in in-vehicle non-driving related tasks (NDRT)). Beyond the sole perspective of mitigating motion sickness symptoms, this study comprehensively examines the potential negative impacts of visual cues in real-world scenarios, providing insights and theoretical references for further research in related fields.
It is important to note that this study has several experimental limitations, resulting in data integrity and reliability that have not yet met the standards for formal publication:
\begin{enumerate}
    \item Small sample sizes (15 participants, 10 participants) due to limited research funding and experimental conditions, which may affect statistical power and the generalizability of results;
    \item Room for optimization in experimental methodology, such as unclear counterbalancing of the order of multiple experimental conditions, lack of justification for the rationality of the 20-minute interval, and absence of sample size power analysis;
    \item Data limitations: the experimental design lacks objective attention metrics (e.g., eye-tracking data), and statistical methods have flaws. As a result, the core hypothesis has not been fully validated, leading to insufficient feasibility of the conclusions.
\end{enumerate}
Despite the aforementioned limitations, the core scientific idea and innovative concept of this study still hold exploratory value. We release this preprint to share the research framework, provide references for peers, and facilitate discussions on experimental optimization. Future work will further validate the hypothesis by expanding sample sizes, optimizing experimental methods, and improving the validation system. The refined results will be separately submitted for formal publication.
We appreciate the attention and constructive comments from fellow researchers, and welcome discussions on the research idea and directions for experimental optimization.

\appendix
\section{Experimental validity analysis}
MSDV, formulated in the Chinese standard GB/T13441.1-2007 \citep{ISO1} and the international standard ISO2631 \citep{vongierke1975iso}, assesses MS likelihood based on axis-specific calculations. A higher MSDV suggests increased risk of motion-related nausea. Turner and Griffin \citep{international1997evaluation} validated this model for predicting vehicle-induced MS. In Experiment 1, MSDV showed no notable differences across X-axis (F(2,42)=0.736, p=0.485, \(\eta^{2}\)=0.034), Y-axis (F(2,42)=0.057, p=0.945, \(\eta^{2}\)=0.003), and Z-axis (F(2,42)=0.841, p=0.438, \(\eta^{2}\)=0.039). Experiment 2 results were similar for X-axis (F(3,36)=0.825, p=0.489, \(\eta^{2}\)=0.064), Y-axis (F(3,36)=1.109, p=0.358, \(\eta^{2}\)=0.085), and Z-axis (F(3,36)=1.314, p=0.285, \(\eta^{2}\)=0.099). These findings excluded driver-induced errors in Figure 11.

\begin{figure}[h]
    \centering
    \includegraphics[width=1\linewidth]{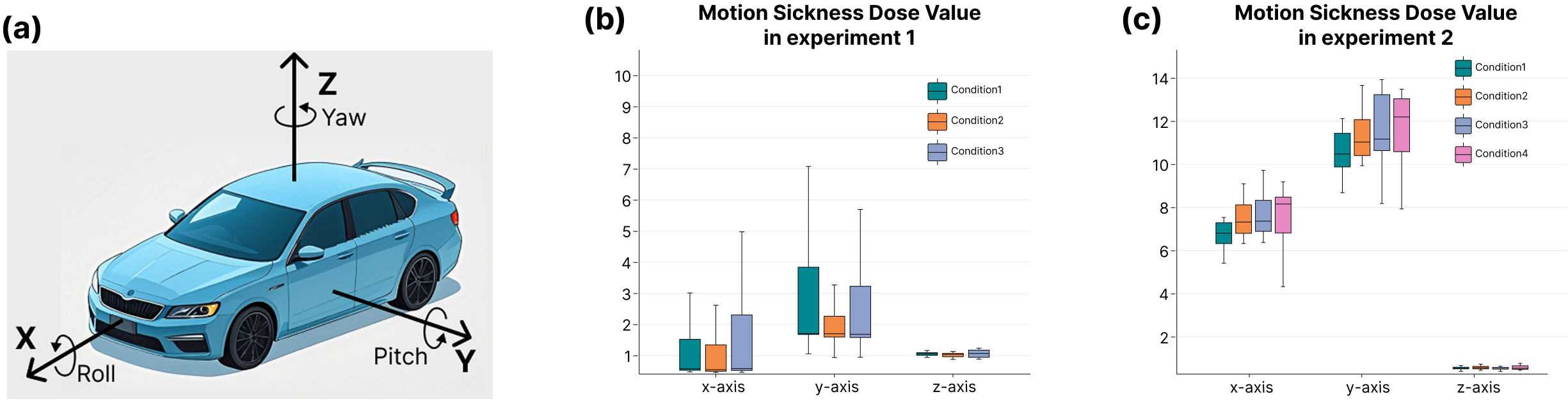}
    \caption{(a). Definition of Vehicle Coordinate Axes; (b). MSDV for each axis across the three conditions in experiment1; (b). MSDV for each axis across the four conditions in experiment2; }
    \label{fig:enter-label}
\end{figure}

Open road experiments face limitations, chiefly the difficulty in maintaining consistent driving patterns across trials. In this research, we used MSDV to confirm no significant disparities in MS impacts from driving patterns across and within groups. However, unpredictable events, like abrupt decelerations caused by other vehicles, can occur, potentially distorting results and data interpretation. Future studies should implement additional strategies to mitigate these issues. By the way, in terms of the analysis of MS susceptibility, in the initial stage of the experiment, two individuals with minimal MS susceptibility were initially chosen for analysis. Results showed their MS levels were nearly nonexistent, markedly differing from those with higher susceptibility. Thus, the final experiment included only those with high MS susceptibility to maintain data validity and consistency.

\section{Dynamic MS perception characteristics of participants}
Dynamically recording passengers' MS levels during extended driving sessions is a crucial element of this study. Data from experiment indicate that sharp turns and sudden decelerations elevate MS levels, while long, straight roads have a calming effect, which aligns with previous studies \citep{guignard1990accelerative,turner1999motion,zhiyong2016adaptive,balk2013simulator}. Figure 12, generated via Python's folium.plugins.Heatmap, displays these changes, with red marking high-frequency MS modifications and green indicating low-frequency ones. 

\begin{figure}[h]
    \centering
    \includegraphics[width=1  \linewidth]{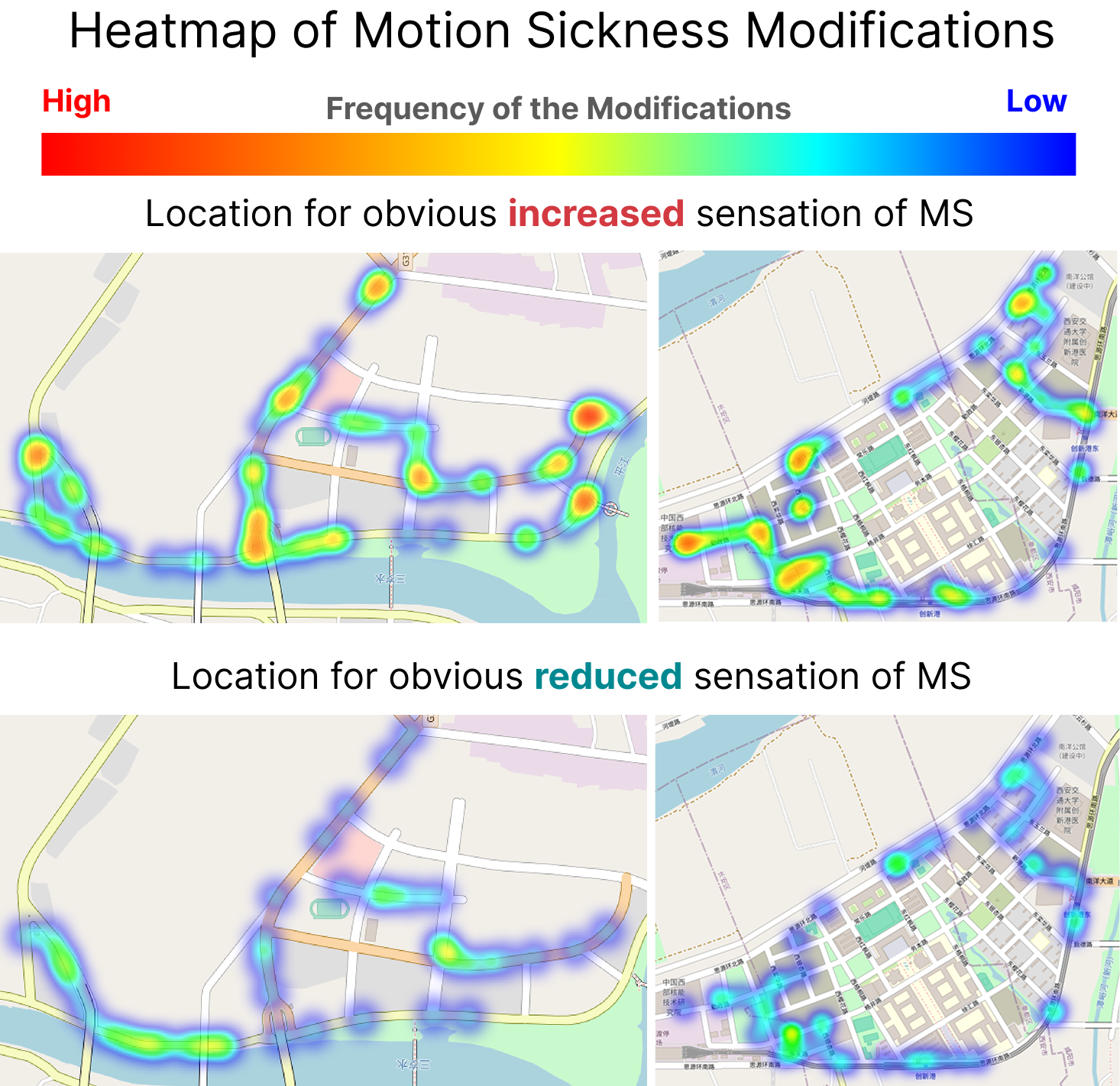}
    \caption{Heatmap of MS Modifications in Participants}
    \label{fig:enter-label}
\end{figure}

Elevated MS levels primarily occur on curves, highlighting them as the main trigger for MS sensation, while straight roads generally reduce MS from a higher level, particularly when there is negligible acceleration or deceleration. This is due to the gradual subsidence of MS symptoms without aggravating influences. Occasionally,  MS deviations appear on curves because some participants mistakenly set the response slider too high. This visual depiction elucidates the relationship between road configurations and MS, providing a foundation for comprehensive analysis of MS in relation to driving conditions.
\printcredits

\bibliographystyle{cas-model2-names} % 简写模式引用文献
% \bibliographystyle{unsrt}  %数字格式引用文献
\bibliography{cas-refs}


% Author biography without author photo.

% \bio{figs/cas-pic1}
% Author biography with author photo.

% \vskip3pc

% \bio{figs/cas-pic1}
% Author biography with author photo.

\end{document}